\documentclass{aa}  
\usepackage{graphicx}
\usepackage{natbib}
\newcommand{\HI}{H{\,\small I}}
\newcommand{\FRI}{FR{-\small I}}
\newcommand{\FRII}{FR{-\small I}}
\usepackage{txfonts}
\begin{document}
\bibliographystyle{aa}
   \title{Large-scale HI in nearby radio galaxies: segregation in neutral gas content with radio source size}

   \subtitle{}

   \author{B.H.C. Emonts\inst{1,2} \and R. Morganti\inst{1,3} \and T.A. Oosterloo\inst{1,3} \and J.M. van der Hulst\inst{1} \and G. van Moorsel\inst{4} \and C.N. Tadhunter\inst{5}. 
          }

   \offprints{B.H.C. Emonts (current address: Columbia University, NY)}

   \institute{Kapteyn Astronomical Institute, University of Groningen, P.O. Box 800, 9700 AV Groningen, the Netherlands
         \and
Department of Astronomy, Columbia University,  Mail Code 5246, 550 West 120th Street, New York, N.Y. 10027, USA\\
              \email{emonts@astro.columbia.edu}
         \and
Netherlands Foundation for Research in Astronomy, Postbus 2, 7990 AA Dwingeloo, the Netherlands
         \and
National Radio Astronomy Observatory, Socorro, NM 87801, USA
         \and
Department of Physics and Astronomy, University of Sheffield, Sheffield S3 7RH, UK
}
   \date{}

% \abstract{}{}{}{}{} 
% 5 {} token are mandatory
 
  \abstract{We present results of a study of neutral hydrogen (\HI) in a complete sample of nearby non-cluster radio galaxies. We find that radio galaxies with large amounts of extended \HI\ ($M_{\rm HI} \gtrsim 10^9 M_{\odot}$) all have a compact radio source. The host galaxies of the more extended radio sources, all of Fanaroff $\&$ Riley type-I, do not contain these amounts of \HI. We discuss several possible explanations for this segregation. The large-scale \HI\ is mainly distributed in disk- and ring-like structures with sizes up to 190 kpc and masses up to $2 \times 10^{10} M_{\odot}$. The formation of these structures could be related to past merger events, although in some cases it may also be consistent with a cold-accretion scenario.

   \keywords{Galaxies: active -- Galaxies: interactions -- (Galaxies:) cooling flows -- Galaxies: ISM -- ISM: jets and outflows}
             }  

\authorrunning{Emonts et al.}
\titlerunning{Large-scale \HI\ in nearby radio galaxies}

   \maketitle
%
%________________________________________________________________

\section{Introduction}

Are different types of radio sources intrinsically different or do they inhabit different galactic environments? We address this long-standing 'nature vs. nurture' issue in a study of the neutral hydrogen (\HI) gas in nearby radio galaxies. \HI\ gas gives insight in the content and distribution of inter-stellar medium (ISM) in the host galaxies of the radio sources. It can be traced both at large scales (in emission), as well as in the central region of the galaxy (in absorption against the radio continuum), where it may serve as fuel for the active galactic nucleus (AGN). \HI\ can also serve as a long-lived tracer of the galaxy's history. For example, large-scale \HI\ structures may have formed from past merger or interaction events \citep[e.g.][]{hib96}, which are often invoked to trigger nuclear activity. 

In this Letter we present and discuss a segregation that we find in large-scale \HI\ content between compact and extended radio sources, the latter all of \citet{fan74} type-I (\FRI). The results are based on a sample of 21 radio galaxies from the B2-catalogue ($F_{\rm 408MHz} \ga 0.2$ Jy) up to a redshift of {\sl z} $\approx$ 0.04. This sample is {\sl complete}, with the restriction that we left out sources in dense cluster environments (since here large-scale gaseous features are likely wiped out on relatively short timescales) and BL-Lac objects. In addition we observed NGC 3894, which has a compact radio source with radio power comparable to our B2-sample sources. In total we observed 9 compact ($< 15$ kpc) and 13 extended \FRI\ ($> 15$ kpc) radio sources. The sources have a radio power 22.0 $\leq$ Log ($P_{\rm 1.4\ GHz}$) $\leq$ 24.8 with no bias in $P_{\rm 1.4\ GHz}$ between the compact and extended sources. The radio sources are hosted by {\sl early-type galaxies (E and S0)}. Observations were made during various observing runs in the period Nov. 2002 - Feb. 2005 with the Very Large Array (VLA) in C-configuration and the Westerbork Synthesis Radio Telescope (WSRT). A full description of the sample and observing details will be presented in a future paper (Paper II).

\section{Results}
\label{sec:results}

\begin{figure*}
\centering
\includegraphics[width=14.6cm]{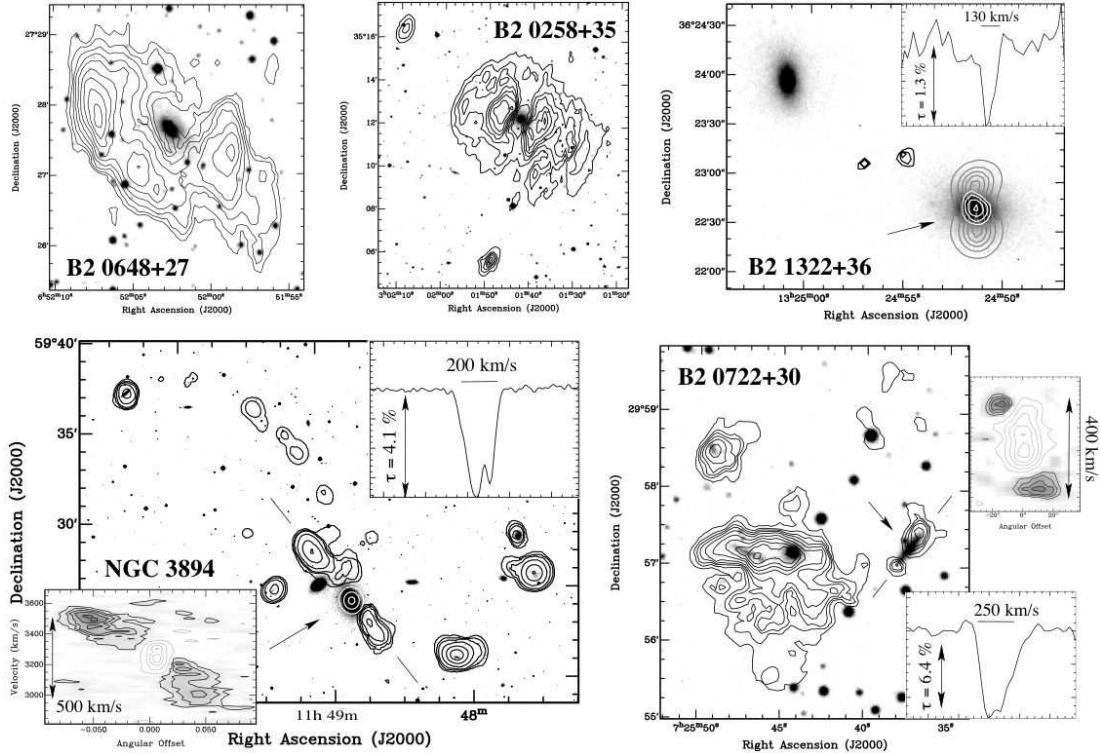}
\caption{0th-moment total intensity maps of the \HI\ emission (contours) around five nearby radio galaxies. Radio continuum is only shown for B2 1322+36 (grey contours); for the other sources, the radio continuum is unresolved (or only marginally resolved for B2 0722+30). Although \HI\ absorption is present in all five radio galaxies, we only show the \HI\ absorption (white contours/profile) in case it clarifies the morphology of the large-scale \HI. The arrows mark the host galaxies of our sample sources, while the broken lines show the direction along which the position-velocity (PV) plots are taken. B2 0648+27: contour levels 0.22, 0.36, 0.52, 0.71, 0.95, 1.2, 1.5, 1.8, 2.1 $\times 10^{20}$ cm$^{-2}$ \citep[see also][]{emo06}; B2 0258+35: levels from 0.34 to 3.0 in steps of 0.44 $\times 10^{20}$ cm$^{-2}$; B2 1322+36: 1.7, 2.3, 2.8 (black) $\times 10^{20}$ cm$^{-2}$; absorption (white) at 17, 45, 70, 95$\%$ of peak absorption -- continuum levels: from 22 to 200 in steps of 44.5 mJy beam$^{-1}$ (grey); NGC 3894: levels 0.17, 0.49, 0.87, 1.7, 3.2, 4.6 (black) $\times 10^{20}$ cm$^{-2}$; absorption (white) at 4, 43, 87$\%$ of peak absorption (PV: levels -1.0, -5.0, -10, -14 (grey), 1.0, 2.0, 3.0, 4.5, 6.5 (black) mJy beam$^{-1}$); B2 0722+30: levels 0.67, 1.3, 1.8, 2.3, 3.0, 4.0, 5.0, 6.0, 7.0, 8.0 $\times 10^{20}$ cm$^{-2}$ -- part of the \HI\ disk that is observed in absorption is not plotted -- (PV plot: levels -0.5, -1.4, -2.4, -3.4, -4.4 (grey), 0.5, 0.7, 0.9, 1.1, 1.3 (black) mJy beam$^{-1}$).}
\label{fig:hi}
\end{figure*}

We detect \HI\ in emission associated with 6 of our objects. The total intensity images and properties of the \HI\ structures (if not previously published) are shown in Figure \ref{fig:hi} and Table \ref{tab:hiradiogalaxies}.\footnote{We assume H$_{0}$ = 71 km s$^{-1}$ Mpc$^{-1}$ and use the redshift (unless otherwise indicated) to calculate the \HI\ properties.} In most cases the \HI\ is distributed in a fairly regularly rotating disk or ring, although a varying degree of asymmetry is still visible in these structures. For B2 0258+35, B2 0648+27 and NGC 3894 the \HI\ structure is more than 100 kpc in size and contains at least about the \HI\ mass of the Milky Way.\footnote{$M_{\rm HI,\ \rm MW} = 5 \times 10^9 M_{\odot}$ \citep{hen82}.} For B2 0722+30 part of the \HI\ disk is seen in absorption against the radio continuum, therefore the total mass of the \HI\ disk is somewhat larger than given in Table \ref{tab:hiradiogalaxies}. The environment of B2 0722+30 (see also Fig. \ref{fig:hi}) is very dynamic and gas-rich (the total mass in the field of B2 0722+30 is $M_{\rm HI} \approx 4 \times 10^{9} M_{\odot}$). B2 1322+36 shows ``blobs'' of \HI\ emission, as well as slightly extended \HI\ absorption, in the direction of a nearby companion, possibly tracing a bridge- or tail-like structure. Although \HI\ absorption is detected in the central beam of all the radio galaxies in Fig. \ref{fig:hi}, as well as a few extended \FRI\ sources in our sample (which we will present in Paper II), this Letter is restricted to the \HI\ gas detected in {\sl emission} in large-scale structures that stretch beyond the optical extent of the host galaxy.
\begin{table}
\caption{{\sl \HI\ in radio galaxies.} Given is the B2 name, the NGC number, the total \HI\ mass detected in emission, the diameter of the \HI\ structure (or distance to the host galaxy for B2 1322+36), the peak in \HI\ surface density, and the morphology of the \HI\ structure (D = disk, R = ring, B = ``blob'').}
\label{tab:hiradiogalaxies}
\begin{tabular}{llccccc}
$\#$ & B2 Name & NGC & {\sl M}$_{\rm HI}$ & {\sl D}$_{\rm HI}$ & $\Sigma_{\rm HI}$ & Mor. \\
 &  &  & ({\sl M}$_{\odot}$) & (kpc) &({\sl M}$_{\odot}$/pc$^{2}$) & \HI \\
\hline
\hline
1 & 0258+35         & 1167 &  1.8$\times$10$^{10}$ & 160 & 2.7  & D \\
2 & 0648+27$^{\ a}$ & -     &  8.5$\times$10$^{9}$  & 190 & 1.7 & R \\
3 & 0722+30         & -     &  2.3$\times$10$^{8}$  & 15  & 4.1 & D \\
4 & 1217+29$^{\ b}$ & 4278 &  6.9$\times$10$^{8}$  & 37 & - & D \\
5 & 1322+36         & 5141  &  6.9$\times$10$^{7}$  & 20  & 3.7 & B \\
6 & -               & 3894  &  2.2$\times$10$^{9}$  & 105 & 3.8 & R \\
\hline
\hline
\end{tabular}\\
\vspace{1mm} 
References: a). \citet{emo06}; b). \citet{mor06b}.
\end{table}

\begin{figure}
\centering
\includegraphics[width=7.9cm]{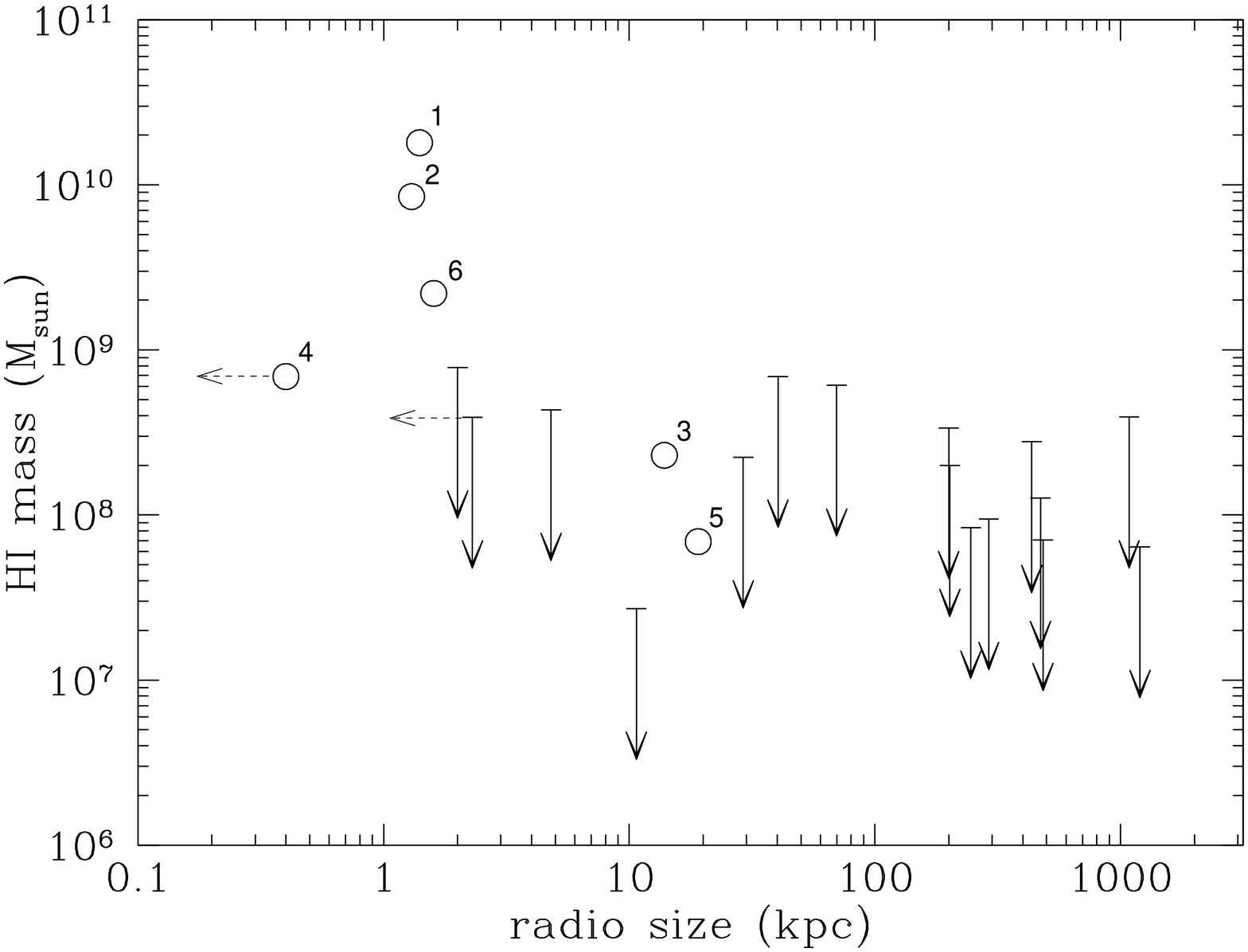}
\caption{Total \HI\ mass detected in emission plotted against the total linear extent of the radio sources. Source sizes have been taken from numerous publications \citep[see][and references therein]{emo06thesis}. In case of non-detection a tight upper limit (3$\sigma$ across 400 km s$^{-1}$) is given. The numbers correspond to the sources as they are given in Table \ref{tab:hiradiogalaxies}.}
\label{fig:hisizeplot}
\end{figure}

The most intriguing result from this Letter is shown in Figure \ref{fig:hisizeplot}. Large amounts of extended \HI\ (with $M_{\rm HI} \gtrsim 10^{9} M_{\odot}$) are only observed around host galaxies with a {\sl compact} radio source, while none of the host galaxies of the more extended FR-I type radio sources shows similar amounts of large-scale \HI. For the non-detections we derived a firm upper limit by estimating the total \HI\ mass of a potential 3$\sigma$ detection smoothed across a velocity range of 400 km s$^{-1}$ (similar to that of the detected \HI\ disks/rings). Note that this does not exclude the possibility that the extended \FRI\ sources contain more conservative amounts of large-scale \HI. Take for example the very nearby radio source Centaurus A, which has a total linear extent of 650 kpc. Cen~A contains $1.5 \times 10^{8} M_{\odot}$ of \HI\ in diffuse shells outside the optical host galaxy \citep{gor90,sch94}. In addition, Cen~A has an inner disk of $M_{\rm HI} \approx 4.5 \times 10^{8}$, but this disk would be only marginally resolved at the distance of our sample sources and resolution of our observations and it would appear mostly in absorption against the central radio continuum.

When applying a statistical Mann-Whitney U-Test on our data, the group of compact ($<$15 kpc) radio sources differs in large-scale \HI\ mass content from the group of extended ($>$15 kpc) radio sources already at the 95$\%$ significance level (even when leaving NGC 3894 out of the statistics). This suggests that for these nearby radio galaxies there indeed is {\sl a segregation in large-scale neutral hydrogen content between compact and extended radio sources}.

\section{Discussion}
\label{sec:discussion} 

The question arises what could cause this segregation in \HI-content between the compact radio sources and the extended \FRI\ radio sources? To answer this question one needs to understand more about the nature - and therefore about the origin - of these large-scale \HI\ structures. One possibility is that they originated from merger events. In a major merger between gas-rich galaxies \citep[e.g.][]{hib96,mih96,bar02}, part of the gas is transported to the central region, while another part is expelled in large structures of low surface brightness (tidal-tails, bridges, shells, etc.). If the environment is not too hostile and the gas in these large structures remains gravitationally bound to the system, it can fall back onto the galaxy and settle into a disk- or ring-like structure within a few galactic orbits \citep[$>$1 Gyr;][]{bar02}. For B2 0648+27 we confirmed a merger origin through the detection of a post-starburst stellar population \citep{emo06}. Another possible formation mechanism of the large-scale \HI\ structures is cold accretion. Simulations by \citet{ker05} show that galaxies can accrete gas from the inter-galactic medium via a cold mode, i.e. the gas cools along filamentary structures without being shock-heated. On smaller scales, \citet{kau06} show that through the cooling of hot halo gas, cold gas can be assembled onto a galactic disc.\\
\vspace{-2mm}\\
Leaving open the exact formation mechanism, we discuss several possible explanations for the segregation in large-scale \HI-content between the compact sources and the extended \FRI\ sources:\\
\vspace{-2mm}\\
The first possibility is that the large-scale \HI\ content and type of radio source both depend on the properties of the host galaxy. As we will describe in Paper II, we find that, using the available optical information, no other properties of the host galaxy correlate strongly with the presence of extended \HI. However, deep optical imaging, as well as information at other wavelengths, is necessary to study in more detail the properties of the host galaxies.\\
\vspace{-2mm}\\
The second possibility is that the radio sources heat/ionise the gas when jets/lobes propagate outward. An example of this is seen in the radio galaxy Coma A, where an \HI\ disk (60 kpc in diameter) is partly ionised by radio jets/lobes expanding into it \citep{mor02}. Simulations of powerful radio sources by \citet{bic06} show that propagating radio jets can create a quasi-spherical, high pressured bubble of radio plasma, that drives radiative shocks in every direction through the host galaxy's ISM. This mechanism - if important also for \FRI\ sources - might explain how extended neutral gas structures (such as the ones in Fig. \ref{fig:hi}) may have been ionised by the extended radio sources in our sample. Alternatively, \citet{bes06} argue that moderately powerful radio sources can self-regulate the balance between cooling and heating of hot gas surrounding these systems. In light of the cold accretion scenario as a possible formation mechanism of large-scale \HI\ structures, this suggests that propagating radio jets may prohibit \HI\ structures from forming in the first place.

Extensive emission-line studies \citep{bau88,bau89,bau92} show that large-scale features of warm gas are common in nearby powerful radio galaxies, but for \FRI\ sources they are not as extended, massive and regularly rotating as the \HI\ disks/rings that we find around some compact sources. Therefore, we do not expect that the extended sources in our sample have converted similar \HI\ structures into warm emission-line gas. On the other hand, X-ray studies are necessary to investigate the presence of hot gas around our sample sources.\\
\vspace{-2mm}\\
The third possibility is that the \HI-rich compact radio sources do not grow into extended sources, either because they are frustrated by the ISM in the central region of the galaxy, or because the fuelling stops before the sources can expand. If our \HI-rich radio galaxies formed through a major merger of gas-rich galaxies, part of the gas is expelled in large-scale tidal structures, while another part is transported into the central kpc-scale region \citep[e.g.][]{bar02}. The latter could be responsible for frustrating the radio jets if they are not too powerful. Alternatively, while the geometry and the conditions of the encounters appear to be able to form the observed large-scale HI structures, they might not be efficient in channelling substantial amounts of gas to the the very inner pc-scale region. This might prevent stable fuelling of the AGN and hence large-scale radio structures do not develop.

The four most compact and most \HI-rich radio sources in our sample have been studied in detail by \citet{gir05a,gir05b} and \citet{tay98}. Interaction with the ambient medium has been suggested for all these sources. Although it is not clear how much gas is needed to confine a radio source \citep[see e.g. the discussion by][]{hol03}, Giroletti et al. argue that the relatively low power radio sources in NGC 4278 and B2 0648+27 can not bore through the local ISM. B2 0258+35 displays variable levels of activity, suggestive of inefficient fuelling, and is therefore not expected to grow beyond the kpc scale \citep{gir05a}.

The extended \FRI\ sources could be fed in another way. One possibility are cooling flows \citep[e.g.][]{fab94}, which may create a continuous supply of gas, allowing the source to grow. Since only a very small amount of gas is necessary to fuel a radio source \citep[e.g.][]{gor89}, another possibility is that \FRI\ sources are related to dry mergers between gas-poor elliptical galaxies \citep[e.g.][]{col95}.

\subsection{Relation to other studies}
\label{sec:otherstudies} 

In a study of Ly$\alpha$ emission-line halos around radio galaxies at $z>2$, \citet{oji97} detect numerous cases of strong \HI\ absorption in the Ly$\alpha$ emission-line profiles, mainly in the smaller radio sources. They prefer the possible explanation that these small radio sources reside in dense, possibly (proto) cluster environments, where the radio source vigorously interacts with the ambient medium. It is worth noting that our nearby radio galaxies were selected {\sl not} to lie in dense clusters. \citet{vil03} argue that Ly$\alpha$ halos at $z>2$ might be cooling flow halos that could be related (progenitors?) to large-scale \HI\ structures at low-$z$. A detailed comparison between low and high redshift \HI-rich radio galaxies deserves further attention.

Finally, large-scale \HI\ is also found around the compact radio source PKS 1718-649 \citep{ver95}, as well as around radio-quiet early-type galaxies \citep[e.g.][]{gor97,mor97,mor06b,oos02,ser06}. Despite the small number statistics and difference in observational restrictions, so far no major differences in \HI\ characteristics (detection rate, morphology and mass) appear between samples of radio-quiet and our sample of radio-loud early-type galaxies \citep[see][]{emo06a}. This might indicate that the radio-loud phase (although not necessarily at the level of the more powerful \FRII\ sources) is just a short period in the lifetime of many, or maybe even all, early-type galaxies.

\section{Conclusions}
\label{sec:conclusions}

We find a segregation in large-scale neutral hydrogen content with radio source size in a complete sample of nearby non-cluster radio galaxies. Large-scale \HI\ structures that stretch beyond the optical host galaxy and that have $M_{\rm HI} \gtrsim 10^9 M_{\odot}$ detected in emission are found only around galaxies with a compact radio source, while the more extended \FRI\ radio galaxies do not show similar amounts of large-scale \HI. If confirmed by studies of larger samples, the large-scale neutral hydrogen content may be a specific property of the host galaxy for various types of radio sources throughout the universe.

\begin{acknowledgements}
We would like to thank J. van Gorkom for useful discussion and the referee for the detailed comments that improved this paper. Part of this project is funded by the Netherlands Organisation for Scientific Research (NWO) under Rubicon grant 680.50.0508. The National Radio Astronomy Observatory is a facility of the National Science Foundation operated under cooperative agreement by Associated Universities, Inc. The Westerbork Synthesis Radio Telescope is operated by the Netherlands Foundation for Research in Astronomy (ASTRON) with support from NWO.
\end{acknowledgements}

\bibliography{emonts_bib_jan07.bib}

\end{document}